# Accepted Manuscript

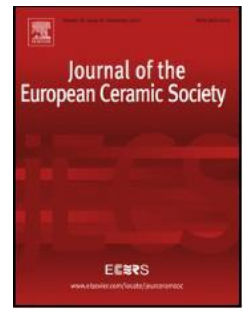

Title: Three-dimensional printed Yttria-stabilized Zirconia self-supported electrolytes for Solid Oxide Fuel Cell applications

Authors: S. Masciandaro, M. Torrell, P. Leone, A. Tarancon

PII: S0955-2219(17)30772-0
DOI: https://doi.org/10.1016/j.jeurceramsoc.2017.11.033
Reference: JECS 11579

To appear in: Journal of the European Ceramic Society

Received date: 4-8-2017
Revised date: 20-10-2017
Accepted date: 13-11-2017

Please cite this article as: Masciandaro S, Torrell M, Leone P, Tarancon A.Three-dimensional printed Yttria-stabilized Zirconia self-supported electrolytes for Solid Oxide Fuel Cell applications.Journal of The European Ceramic Society https://doi.org/10.1016/j.jeurceramsoc.2017.11.033

**Three-dimensional printed Yttria-stabilized Zirconia self-supported electrolytes for Solid Oxide Fuel Cell applications**

S. Masciandaro[a,b], M. Torrell[a], P. Leone[b], A.Tarancon[*,a]

[a] IREC, Catalonia Institute for Energy Research. Jardins de les Dones de Negre 1. PL2, 08930 Sant Adrià de Besòs, Barcelona, Spain, atarancon@irec.cat

[b] PoliTo, Politecnico di Torino, Corso Duca degli Abruzzi, 24- 10129 Torino, Italy

**Abstract**

Additive manufacturing represents a revolution due to its unique capabilities for freeform fabrication of near net shapes with strong reduction of waste material and capital cost. These unfair advantages are especially relevant for expensive and energy-demanding manufacturing processes of advanced ceramics such as Yttria-stabilized Zirconia, the state-of-the-art electrolyte in Solid Oxide Fuel Cell applications. In this study, self-supported electrolytes of yttria-stabilized zirconia have been printed by using a stereolithography three-dimensional printer. Printed electrolytes and complete cells fabricated with cathode and anode layers of lanthanum strontium manganite- and nickel oxide-yttria-stabilized zirconia composites, respectively, were electrochemical characterized showing full functionality. In addition, more complex configurations of the electrolyte have been printed yielding an increase of the performance entirely based on geometrical aspects. Complementary, a numerical model has been developed and validated as predictive tool for designing more advanced configurations that will enable highly performing and fully customized devices in the next future.

**Keywords:** Solid Oxide Fuel Cell; 3D printing; Stereolithography; yttria-stabilized zirconia; electrolyte

## 1. Introduction

Advanced manufacturing processes for ceramic Solid Oxide Fuel Cells (SOFCs) have been successfully developed for decades being currently a mature technology [1]-[3]. Their intrinsic limitations for generating complex geometries have restricted the SOFC technology to simple planar and tubular configurations, mainly fabricated by tape casting and extrusion techniques, respectively [3]. In addition, the so-fabricated individual SOFC cells have to be integrated in complete stacks finally involving a huge number of processing steps, typically including manual assembly and multiple joints [5], which reduces the reliability of the process, increases the complexity of the design for manufacturing and entails longer time-to-market. Taking this into consideration, the high level of geometry complexity allowed by three-dimensional (3D) printing technologies could represent a major revolution in the design and fabrication of SOFCs and SOFC stacks. In this sense, one can easily anticipate the unfair advantages arising from exploring unconventional monolithic configurations in terms of customization and integration of SOFCs into power generation systems. Moreover, the monolithic integration of a full-ceramic freeform SOFC stack could eventually suppress critical joints therefore representing a step change in its reliability and durability. The application of 3D printing technologies to the fabrication of SOFCs would also bring a simplification of the manufacturing process, a reduction in the number of fabrication steps and a strong decrease of the initial investment and production costs, which is considered one of the main barriers for the market entrance of the SOFC technology [6].

Accordingly, the use of 3D printing technologies for the fabrication of multi-material functional ceramics like SOFCs is highly desirable and advantageous. However, whilst 3D printing of polymeric and metallic objects is being used, even industrially, for already a decade [7]-[10], the fabrication of 3D printed ceramic parts is still mainly under research level development [11]. Among other, 3D printing technologies including selective laser sintering (SLS) [13], direct [14] and indirect [15] inkjet printing (DIP and IIP) and stereolithography (SLA) [16] have been used so far for the fabrication of functional ceramic devices. In particular, the reader is referred to the review by Ruiz-Morales et al. [12] on the most recent developments of 3D printing of ceramics in the field of energy and, more specifically, for Solid Oxide Fuel Cells applications. As compiled by Ruiz-Morales et al.[12], a series of pioneering works focused on individually printing functional ceramics of interest for their particular application in SOFCs have been recently published [17]-[27]. Regarding the electrolyte, several of these studies have been devoted to the fabrication of dense yttria-stabilized zirconia (YSZ) layers with a thickness below 10 µm onto different porous and dense substrates. Li et al. [17], Tomov et al. [18], and Sukeshini et al. [19] deposited fully functional YSZ electrolytes by inkjet printing directly on tape cast and pre-sintered NiO-YSZ cermet substrates typically used in anode-supported SOFCs. Full SOFCs based on these inkjet-printed dense electrolytes were fabricated in all cases showing excellent open circuit voltages (OCVs), OCV = 1.1, 1.01 and 1.05 V, respectively, operating under hydrogen at 800ºC. These values approaching the expected theoretical OCVs indicate that the printed electrolytes were gas-tight suitable for producing high-power SOFCs. In the same direction, Esposito et al. [20] were able to deposit gas-tight 1.2 µm-thick YSZ electrolyte layers on large area (~16 cm 2) NiO-YSZ tape cast anode supports by using inkjet printing of highly diluted inks of nanometric ceramic powders obtaining OCV values of 1.07-1.15 V and remarkable peak power density (PPD) of 1.5 W/cm$^2$ at 800ºC under hydrogen atmosphere. Most interestingly, strikingly thin 150 nm-thick YSZ layers were recently obtained by Gadea et al. [21] by using reactive inkjet printing compatible with multiple cost-effective substrates such plastics or metals. This substrateflexibility together with the low temperature calcination treatments (400‑500° C) required for densifying these ultrathin layers open the possibility of integrating them into innovative configurations such as micro-SOFCs [28][29][30].

Despite the excellent results for YSZ layers printed onto different substrates, to the best of the authors' knowledge, there are not references of 3D printed self-supported fully functional YSZ electrolytes for SOFC applications. In this sense, the ability of printing high aspect ratio YSZ parts combined with functional self-supported dense layers is considered crucial for the development of monolithic SOFC devices [12]. In this study, joint-less 3 mol % Yttria-stabilized Zirconia (3YSZ) membranes were printed by stereolithography [31] for being employed as electrolytes in SOFCs. Although 3YSZ presents certain limitations due to their low ionic conductivity (compared to the State-of-the-Art, SoA, 8 mol% YSZ), their excellent mechanical properties make it more interesting for proof-of-concept purposes. Two different 3YSZ membrane geometries were proposed, with planar and honeycomb structures, and implemented in electrolyte-supported cells using YSZ composites with lanthanum strontium manganite (LSM) as a cathode and nickel composites as an anode, i.e. Ni-YSZ/3YSZ/LSM-YSZ. The devices were electrochemically tested at 800, 850 and 900°C and the results were compared with a numerical model able to predict the experimental measurements as a function of the printed geometry.

## 2. Experimental and physical modeling

### 2.1. Fabrication of the 3D printed cells

The fabrication of the ceramic 3YSZ pieces by SLA involved different steps, namely, digital design, printing, cleaning, debinding and sintering. Starting from commercial 3YSZ pastes from 3DCeram, different objects were printed using a CERMAKER SLA printing machine. The paste involved in this process is the commercial 3DMIX-3YSZ composed by a ceramic powder, a UV curable monomer, a photoinitiator and a dispersant, which allows reaching good homogeneity of the powders and a low viscosity of the suspension which is further improved by adding diluents [31][32].

The printing process is based on the doctor blade method. The ceramic paste is spread on the printing platform by using two blades calibrated at 50 µm and 150 µm, being able to scrap thin and uniform layers up to 25 microns. This doctor blade step is followed by photocuring of the UV monomers (acrylate type) using a power modulated UV laser with a wavelength of 355 nm. This photocuring step allows defining a certain pattern with lateral and vertical resolutions of *ca.* 30 µm and 50 µm, respectively. After this, the printing platform moves down 25 µm to fully cover the printing area with a new layer of uncured paste. The whole sequence is repeated as many times as needed. The average printing speed is 100 layers/h, which results in a thickness of ~2.5mm/h for this high-resolution example.

After printing, the uncured paste was removed by using a wax-based cleaning process. Afterwards, clean pieces were thermally treated for more than 70 h in order to eliminate the organics without crack formation on the green parts and, finally, were sintered at 1450ºC for 2 hours. Two different geometries based on flat and honeycomb-like self-supported membranes coupled to bulky rings were fabricated. The flat membrane had a thickness of 340 µm and an active area of 1.54 cm2 . The honeycomb cell (HC) consisted of 260 µm-thick hexagonal cells of ca. 1mm2 forming a network connected by 530 µm-thick beams of 220 µm in width**.** The sintered pieces of 3YSZ were painted with commercial NiO-YSZ and LSM-YSZ pastes as electrode materials. Symmetrical cells with the same electrode material in both sides as well as complete fuel cells with NiO-YSZ as an anode in one side and YSZ-LSM as a cathode material in the other side (NiO-YSZ/YSZ/YSZ-LSM) were fabricated. Different thermal treatments for improving the attachment of the electrode materials were evaluated to minimize polarization losses. Finally, the NiO-YSZ layers were treated in air at 1450ºC for 2h while the YSZ-LSM layers at 1100ºC for 2 h.

## 2.2. Microstructural and electrochemical characterization of the cells

Microstructural and morphological characterization of the sintered pieces of 3YSZ has been performed by employing X-ray Diffraction (XRD) in a Bruker D8 Advance diffractometer in 2θ mode from 20 to 80º, using copper $K_{\alpha 1}$ radiation; optical microscopy in a SENSOFAR microscope; and Secondary Electron Microscopy (SEM) coupled to BackScattered Electrons (BSE) and Energy-Dispersive X-ray spectroscopy (EDX) detectors in a Zeiss Auriga microscope.

A current collector based on platinum mesh and nickel paste was employed in combination with the NiO-YSZ electrode while a platinum mesh and commercial $La_{0.5}Sr_{0.5}CoO_3$ paste (Kceracell) was employed in combination with the YSZ-LSM electrode. In fuel cell mode, *Ceramabond*™ sealant paste was applied to ensure gas tightness between the anode and cathode chambers. This step was less critical in this work since the ring support monolithically integrated in the membrane avoided any gas leakage close to the active electrodes. The electrochemical and fuel cell tests were carried out in a commercial ProboStat™ (NorECs AS) sample holder placed inside a high temperature tubular furnace. The cells were electrically characterized at temperatures from 350 to 900 ºC by employing a multimeter (Keithley Model 2400) and a frequency response analyzer (Novocontrol Alpha-A). Electrochemical Impedance Spectroscopy (EIS) was carried out in a range of frequencies from 5 MHz to 100 mHz and applying an AC signal of 50 mV of voltage amplitude under OCV conditions. In order to obtain a more complete electrochemical characterization of the cells EIS measurements were performed under a DC bias voltage of 0.7 V by employing a potentiostat/galvanostat (Parstat 2273 from PAR). Synthetic air was used as oxidant while pure humidified hydrogen was used as a fuel.

**2.3. Solid Oxide Fuel Cell numerical model**

Numerical models have been implemented using Comsol Multiphysics 5.2a to evaluate the performance of the fabricated Solid Oxide Fuel Cells. Regarding the numerical domain, a two-dimensional axisymmetric geometry with a concentric feeding channel system has been implemented for the flat SOFC configuration, as suggested by Santarelli et al. [38]. This approximation allows obtaining a 3D geometry by applying a revolution through the Z axis of symmetry, taking advantage of a shorter computational time with respect to a complete 3D model. Complementary, a three-dimensional domain has been used for the simulation of the honeycomb SOFC. By taking the advantage of geometry and boundary conditions symmetry, a 1/12 of the whole geometry has been selected as computational domain to reduce the simulation computational time.

The elements of the simulated SOFC are made of different solid and gases, namely, i) an electrolyte of 3YSZ; ii) an anode of Ni-YSZ and a cathode of LSM/YSZ; iii) hydrogen humidified with a 3% of $H_2O$ in the anode chamber while a humidified synthetic air in the cathode chamber. The ionic conductivity of the 3YSZ was obtained from experimental measurements included in this work. The electronic and ionic conductivities of the electrodes were extracted from the literature [39]. Porosities for both the anode (Ni/YSZ) and cathode layers (LSM/YSZ) were obtained from SEM studies, while Carman-Kozeny equation was applied to evaluate the permeability, considering the average particle size diameter obtained through SEM analysis. The specific surface area of the Triple Phase Boundaries (TPB) of the composites was determined using the particle coordination number method in binary random packing spheres and the percolation theory, as it is explained in by J.H. Nam et al. [41]. The input values calculated for the specific surface are of the TPB are included in **Table 1**.

The following mathematical and physical models were considered for simulating the behavior of the cells: i) ohm's law for the electronic and ionic charge balance; ii) Butler-Volmer equations for the charge transfer; iii) Maxwell-Stefan diffusion and convection equations for accounting the mass balance in both gas channels and porous electrodes; iv) Navier-Stokes equations to solve the flow distribution in the gas channels; and v) Brinkman equations to solve the flow in the porous gas diffusion electrodes. In order to simplify the model, the following assumptions were considered according to Santarelli and co. [38]: i) isothermal model; ii) steady-state conditions; iii) laminar flow; iv) reactant species as compressible ideal gases; v) continuity and uniformity of the two conductive phases (electronic and ionic) and vi) pure ionic conduction of the electrolyte. The exchange current densities $i_0$, and $i_0$, employed in the Butler-Volmer equations at the different temperatures were obtained from the experimental data of this work as explained by P. Leone et al. [40]. The characteristics of fluid and solid materials and other input parameters employed for the numerical simulations are summarized in **Table 2**. Boundary conditions equivalent to Shi et al [39] were employed to solve the charge transfer, the mass transfer and the flow distributions.

**3. Results and discussion**

Monolithic 3YSZ pieces consisting of a self-standing membrane on top of a ring support were obtained after printing by stereolithography and subsequent cleaning, debinding and sintering steps (inset **Figure 1**). X-ray Diffraction (XRD) patterns of the sintered piece (not presented here) confirmed a single phase crystalline YSZ. Secondary Electron Microscopy (SEM) was employed to characterize the morphology and microstructure of the 3YSZ membrane. According to the top-view SEM image included in **Figure 1a**, a fully dense and homogeneous microstructure is obtained after sintering at 1450ºC with a submicronic average particle size. The cross-section SEM image of the membrane (**Figure 1b and 1c**) also shows a high density and homogeneity with a fairly constant thickness of about 340 μm. This high density is needed in order to ensure the electrolyte gas tightness required in SOFC applications and to minimize the contribution of the electrolyte to the total resistance of the cell. Punctual defects can be observed in the SEM images (in the form of darker particles). Further analysis of these defects based on Back-Scattered Electrons (BSE) and Energy-Dispersive X-ray spectroscopy (EDX) indicates that they are particles made of other materials occasionally deposited in the printer, mainly alumina. Despite the existence of this incidental cross-contamination, a significant effect on the electrochemical properties of the YSZ is not expected due to the low density of defects present in the layer. The high density and the resolution of the YSZ printed by SLA can be observed in the **Figure 1c**.

In order to electrochemically characterize the 3YSZ membrane, symmetrical and complete fuel cells were fabricated based on nanocomposites of commercial NiO-YSZ and LSM-YSZ deposited on both sides as cathode and anode, respectively. **Figure 2a and b** shows cross-section SEM images of the electrodes attached to the 3D printed electrolyte. An adequate intermixing of YSZ and the corresponding active materials is observed in both cases together with a good homogeneity of the grain size distribution all across the layer. A much finer microstructure is clearly observed for the commercial cathode. The optimum attachment temperature of both electrodes was determined, with especial attention to the one of the cathode side, after systematical evaluation of the polarization resistance in symmetrical cell configuration (not presented here for conciseness reasons). Area specific resistances of ASR=0.4 $\Omega cm^2$ and ASR=0.2 $\Omega cm^2$ at 900ºC and activation energies of $E_a$=1.12±0.04 eV and $E_a$=1.43±0.02 eV were obtained for the optimized anode and cathode layers, respectively, being in reasonable agreement with the literature [43] [44]. A detail of the attachment of a typical cathode to the electrolyte layer is presented in **Figure 2c** showing the good connnectivity required to ensure a high performance in SOFC mode.

The performance of the printed 3YSZ electrolyte was electrochemically evaluated using optimized LSM/YSZ electrodes in symmetrical cell configuration. Electrochemical Impedance Spectroscopy (EIS) measurements were carried out at different temperatures ranging from 350 to 900ºC in synthetic air. **Figure 3** shows the Nyquist plot of the typical spectra obtained at high temperatures together with the corresponding equivalent circuit model. Circuit elements values are reported in **Table 3**. By fitting the experimental values to the equivalent circuit, it is possible to obtain the total resistance associated to the ionic diffusion within the electrolyte. **Figure 4** shows the Arrhenius plot generated by the representation of the contribution of the electrolyte to the total resistance of the cell as a function of the inverse temperature. Two different regions with Arrhenius behavior can be observed at low and high temperatures being in good agreement with the literature [33]. Activation energies of 0.78±0.05 eV for T>560 °C and 1.03±0.03 eV for lower temperatures have been fitted, while the ionic conductivity showed values of 0.011 S/cm at 800 °C and 0.022 S/cm at 900 °C. According to previous studies [29-32], this dual behavior can be fully explained by the defect dissociation-migration energy model.

After proving that the printed 3YSZ membrane presents the typical behavior of the same type of layers fabricated by traditional methods, the membrane was evaluated as electrolyte in full cells (Ni-YSZ/YSZ/YSZ-LSM) operating under typical conditions, namely, water saturated pure hydrogen and synthetic air in the anode and cathode sides, respectively. The ring support monolithically fabricated around the membrane keeps the sealing area far from the active part of the cell avoiding gas leakages associated to the assembly procedure. The gas tightness of the printed electrolyte is therefore confirmed through the Open Circuit Voltage measurement, which reached a value of OCV=1.14±0.05V at 800ºC (**Figure 5**), perfectly matching the theoretical value at this temperature ($OCV_{th}$=1.11V). As naturally expected, the value of OCV decreased when increasing the operating temperature indicating an adequate behavior of the cell. **Figure 5** shows the galvanostatic V-I polarization curves for the flat SOFC in the range of temperatures from 800 to 900ºC. A maximum peak power density of 100mW/cm$^2$ was observed at 900 °C corresponding to a current density of ca. 200 mA/cm$^2$ at 0.5V. A perfect ohmic behavior is observed due to the expected dominant role of the contribution of the electrolyte to the total resistance. This point is confirmed by the EIS measurements of the cell carried out at V=0.7V (**Figure 6a**), where the high value of resistance obtained by the cut of the curve with the x-axis is essentially due to the electrolyte. After fitting an equivalent circuit (similar to the one in the **insert of Figure 3**) and representing the serial and the polarization resistances (mainly associated to the electrolyte and electrodes, respectively), it becomes evident that the contribution of the electrolyte to the total resistance limits the performance of the cell even at high temperatures (**Figure 6b**). This is mainly due to the fact that 3YSZ was employed as an electrolyte instead of other better performing ionic conductors, e.g. the SoA 8YSZ. As discussed in the introduction section, 3YSZ was the choice for this proof-of-concept due to their excellent mechanical properties. Considering this, the performance of the cell is the one expected for this combination of materials proving the feasibility of fabricating SOFC self-supported electrolytes by 3D printing.

After showing the proper performance of a conventional geometry such as a flat membrane, a more complex configuration consisting of a honeycomb-like structure was generated by 3D printing and tested. This honeycomb 3D printed membrane presents an aspect ratio of 40, evaluated as the ratio between the diameter and the thickness weighted average. The **inset of the Figure 7a** shows a typical monolithic 3YSZ membrane obtained after printing by stereolithography and subsequent cleaning, debinding and sintering steps. This geometry was especially designed to enable thinner membranes by using mechanically robust beams (shown in the SEM picture included in **Figure 7a**). Complementary, recent works have reported the benefits of incorporating geometrical patterns to increase the volumetric current density especially in the singularities [45][46]. The thickness of the membrane was 260 µm (one fourth of the flat one) inside the hexagonal cells and ca. 530 µm in the structural beams (**Figure 7b**). The **inset of the Figure 7b** shows a SEM cross-section image of the detail of the morphology of the printed object. The layer-by-layer printing method presents a resolution of about 25 µm, both laterally and vertically, in 3YSZ, which indicates the great potential of this technology to substantially increase the level of complexity of traditional SOFC cells and stacks.

Similarly to the flat membrane, a honeycomb electrochemical cell was prepared by attaching two electrodes, namely, NiO-YSZ for the anode and LSM-YSZ for the cathode. EIS and fuel cell measurements were carried out for the honeycomb cell under equivalent operation conditions. **Figure 8** shows the V-I polarization curves of the HC cell at 800, 850 and 900ºC. The close-to-theoretical OCV values confirmed the gas-tightness and full density of the 3D printed electrolyte membranes even when more complex geometries were pursued. The HC configuration presents power peak densities of 115 mW/cm2 at 900°C and 69mW/cm$^2$ at 800°C, which represents an increase of 15% compared to the flat configuration. According to the different contributions to the total resistance represented in **Figure 9** (obtained by deconvolution of the EIS spectra), this enhancement is mainly associated to the reduction in the serial resistances compared to the flat cell, i.e. to the thinner electrolyte.

A numerical model based on the Finite Element Method was employed to determine the effect of the increased level of geometrical complexity in the enhancement of the final performance. After validation of the model by comparison of the experimental and simulated V-I curves for both the flat and the HC cells (**Figure 10a**), the current density and voltage distributions at the electrolyte were analyzed under real operation conditions (**Figure 10b and c**).

While the current density distribution of the flat electrolyte is naturally constant (not presented here), **Figure 10b** shows the map corresponding to the honeycomb structure that presents a mostly uniform density in the thinner parts (hexagonal cells) with a significant decrease associated to the thicker parts (support beams). Despite this natural decrease in the current density for the thicker beams, a certain extension of the current lines goes beyond the flat hexagonal cells (insert **Figure 10b**). Complementary, the voltage distribution map (**Figure 10c**) shows a continuous voltage drop inside the beams confirming the active contribution of these parts to the final performance even being thicker than the rest of the cell. All in all, the simulations confirm that the honeycomb structure positively contributes to enhance the performance of the cell compared to the flat counterpart due to i) enabling a thinner membrane and ii) partly using the area increase associated to the beams.

The validation of the numerical model developed for complex SOFC geometries combined with the high degree of freedom proved for the 3D printing of functional electrolytes based on YSZ enable the aprioristic design of novel high-performing geometries.

## 4. Conclusions

Three-dimensional printing of high aspect ratio self-supported and joint-less membranes of 3YSZ has been proved in this work (for the honeycomb shaped membranes an aspect ratio of 40 has been achieved). The functionality of these membranes for working as electrolytes has been evaluated in terms of their ionic conductivity and gas-tightness (full density) as part of symmetrical electrochemical cells and complete fuel cells. Ionic conductivity values of 0.022 S/cm were obtained at 900ºC in good agreement with previous works reported in the literature. Under fuel cell operation conditions, open circuit voltages of 1.14±0.05V at 800ºC were measured, perfectly matching the theoretical ones, indicating the gas-tightness of the 3D printed membrane. Power peak densities of 100 mW/cm2 were obtained at 900 °C being compatible with the values expected for SOFCs based on 3YSZ electrolytes of the current thickness. Therefore, a proof of concept for 3D printed electrolyte-based Solid oxide Fuel Cells has been achieved.

Additionally, better performing cells were printed taking advantage of the higher level of geometrical complexity allowed by the 3D printing technique. SOFC cells based on thinner honeycomb-like electrolytes were fabricated yielding an enhancement of the power density (PPD of 115mW/cm2 at 900°C) entirely due to geometrical aspects. A numerical model (Finite Element Method) able to take into consideration the geometry of the cells has also been validated. This model can be used as a predictive tool for designing optimum geometry configurations by anticipating their performance before printing them. This aprioristic design of complex systems such as SOFCs will revolutionize the field of energy by opening a new avenue for the customization of systems, the fabrication of joint-less stacks or the increase of the specific power, among other unfair advantages. In order to reach these ambitious goals, further development is required to increase the list of functional printable ceramic materials and the multi-material capabilities of the current 3D printing technologies.

**Acknowledgements**

The authors want to acknowledge the received funds on the frame of the RETOS program (Spanish Ministry of Economy, Industry and Competitiveness) for the 3D-MADE project (ENE2016-74889-C4-1-R) and the CHISTE (Complex Hierarchical Infiltrated Structures for High Temperature Electrolysis) projects (ENE2013-47826-C4-3-R).

# FIGURES

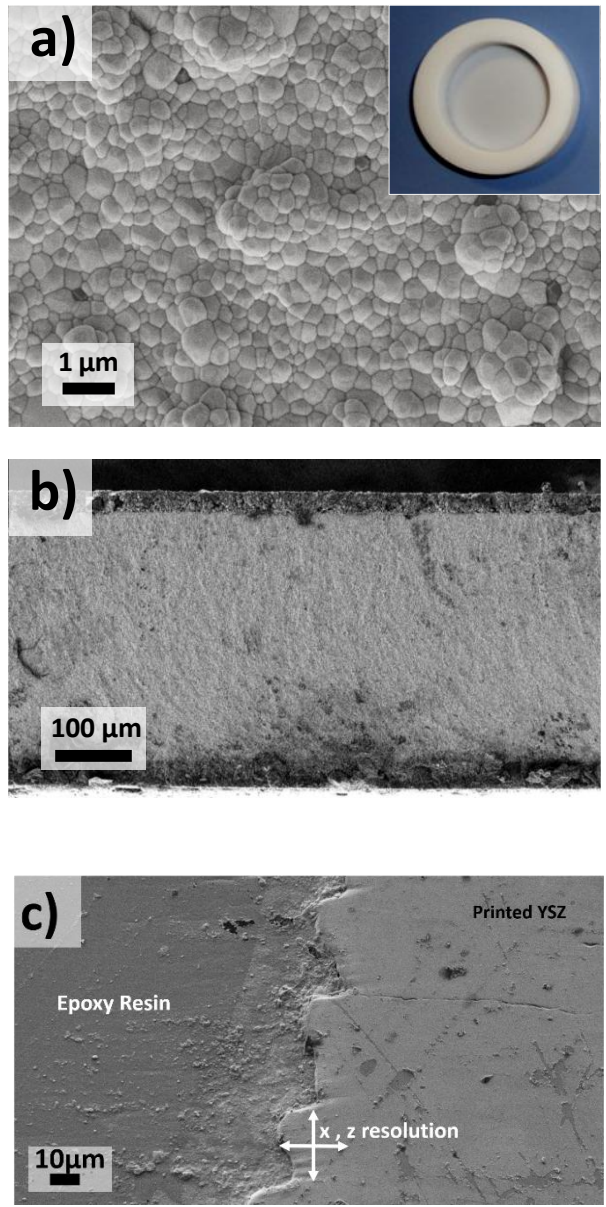

***Figure 1. a)*** *Top-view SEM image of the 3D printed 3YSZ electrolyte. Dark particles corresponding to alumina grains are introduced due to cross contamination. The inset shows an optical image of the self-supported membrane on the bulky ring. SEM images of the cross-section of the self-supported membrane showing **b)** electrodes in both sides and **c)** its high resolution and density.*

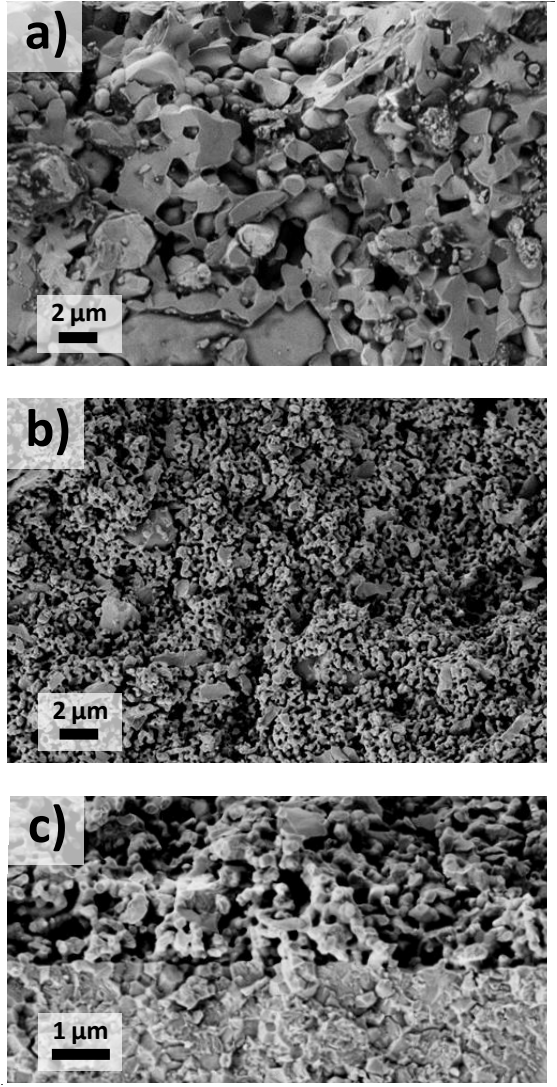

*Figure 2.* Cross-section SEM images of the microstructure of the Ni-YSZ anode (**a**) and LSM-YSZ cathode (**b**); **c**) detail of the attachment of the LSM-YSZ cathode to the 3D printed electrolyte.

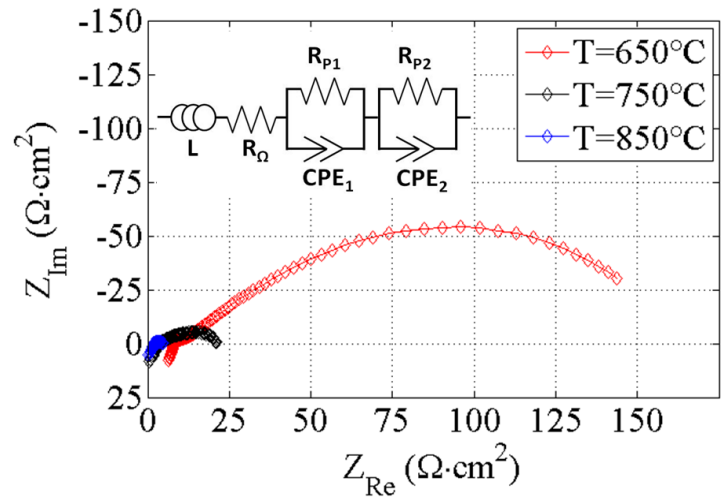

*Figure 3. Nyquist plot of the EIS spectra of the symmetrical LSM-YSZ/YSZ/YSZ-LSM cell measured at different temperatures (T=650-850ºC) in synthetic air. The inset shows the equivalent circuit employed for fitting the arc (solid line).*

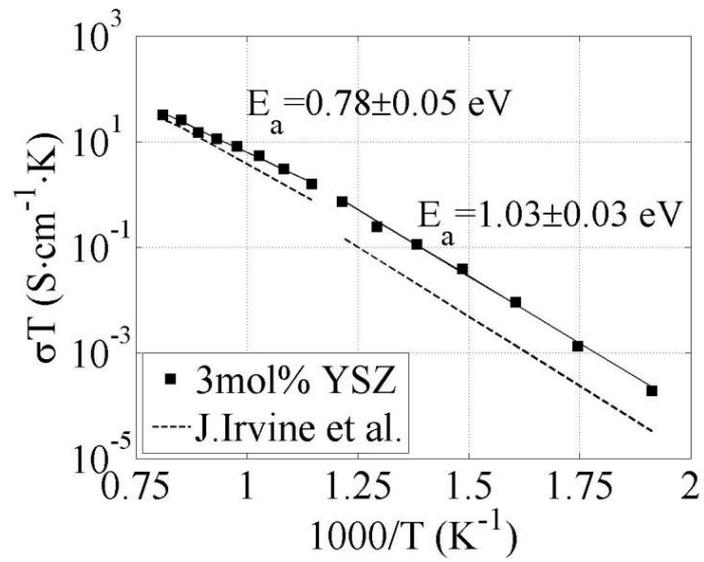

*Figure 4* Arrhenius plot of the ionic conductivity of the 3D printed 3YSZ obtained from the EIS measurements after deconvolution of the spectra. The solid line represents the experimental fitting of an Arrhenius behavior. Data from the literature (J. Irvine et al. in ref. [33]) is included for comparison (dashed line).

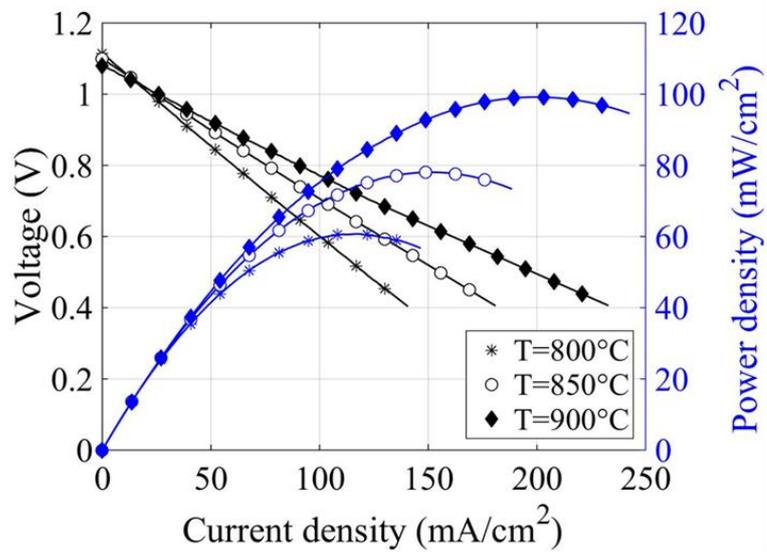

*Figure 5. Voltage- and power- current density curves of the flat fuel cell (Ni-YSZ/YSZ/YSZ-LSM) operated at different temperatures (T=800, 850, 900ºC) under humidified hydrogen and synthetic air atmospheres.*

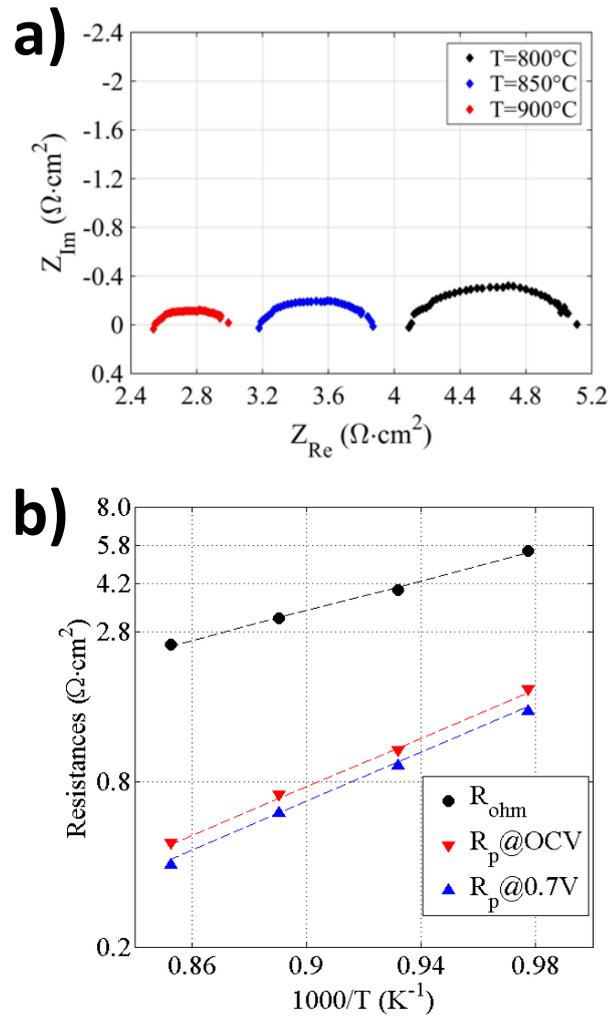

***Figure 6**. **a)** Nyquist plots of the EIS spectra obtained for 3D printed flat fuel cells at V=0.7V at different temperatures (T=800, 850 and 900ºC) under humidified hydrogen and synthetic air conditions; **b)** Ohmic and polarization resistances obtained from deconvolution of the EIS spectra obtained at OCV and V=0.7V as a function of the inverse temperature.*

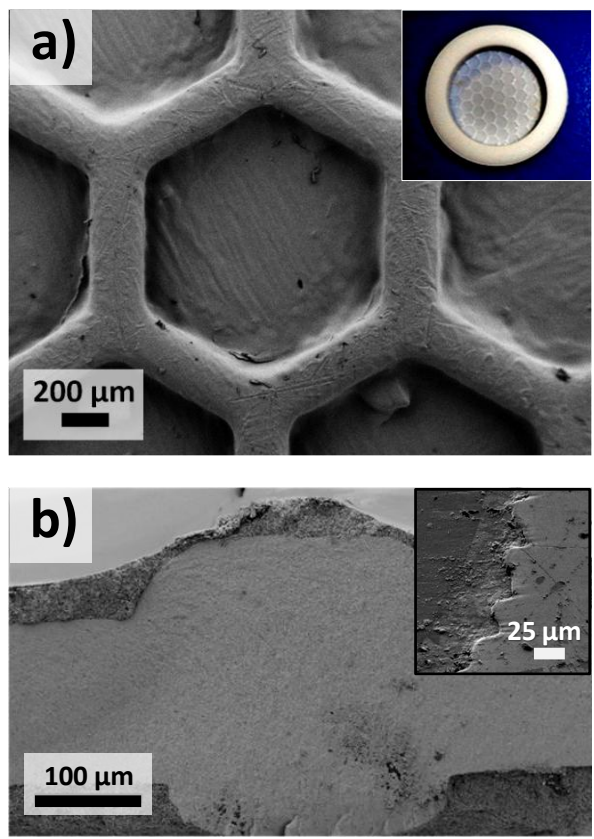

***Figure 7.** a) SEM image of the surface of the 3D printed membrane with honeycomb-like structure. A detail of one of the hexagonal cells and the reinforcing beams can be observed. The inset shows an optical image of the 3D printed piece including the self-supported membrane and the bulky ring; **b**) Cross-section SEM image of one of the beams. The inset shows a detail of the steps generated after each deposited layer in curved geometries. This is an indicator of the resolution of the printing process.*

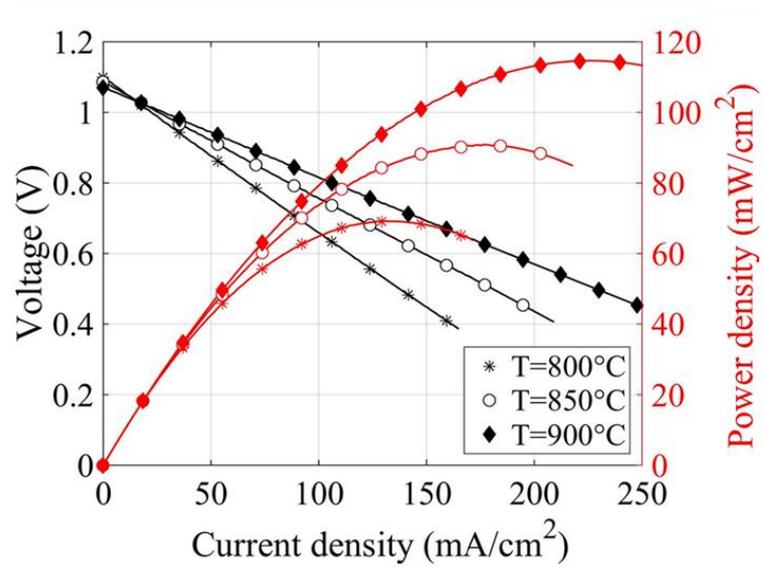

*Figure 8.* Voltage- and power- current density curves of the honeycomb-like electrolyte fuel cell (Ni-YSZ/YSZ/YSZ-LSM) operated at different temperatures (T=800, 850, 900ºC) under humidified hydrogen and synthetic air atmospheres.

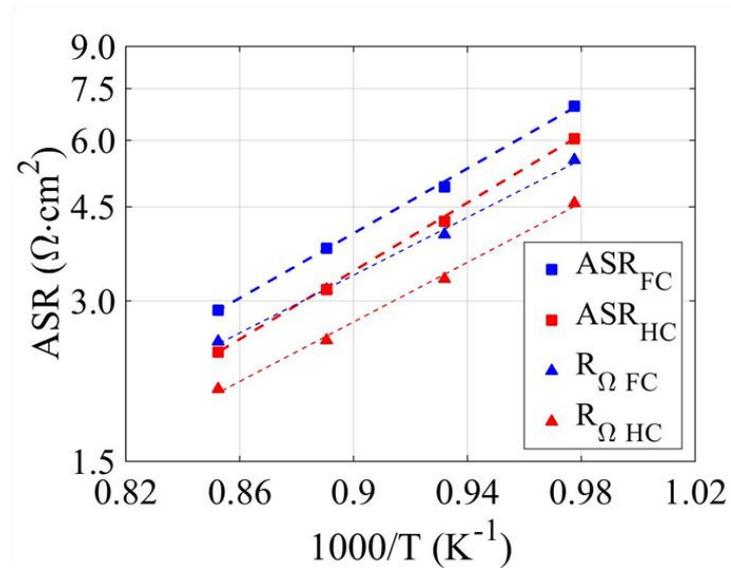

*Figure 9*. *Ohmic and total area specific resistances as a function of the inverse temperature of the flat (FC) and honeycomb-like electrolytes (HC) obtained from deconvolution of the EIS spectra mesured at V=0.7V under fuel cell conditions.*

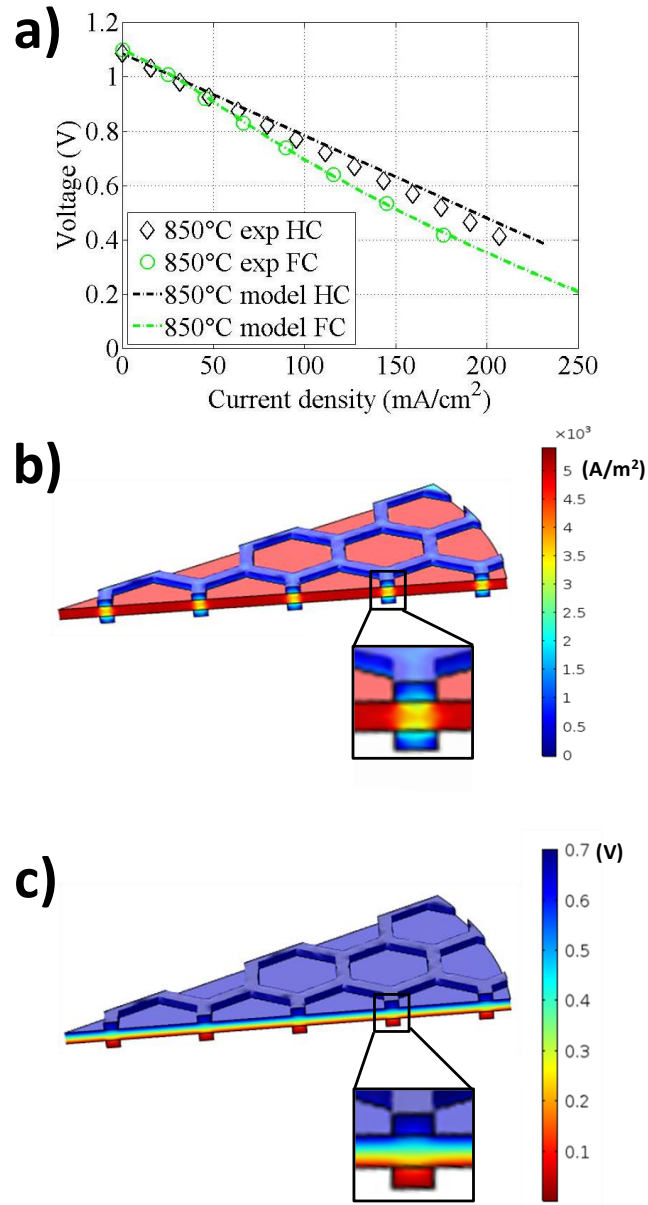

*Figure 10. a)* Voltage-current density curves of flat (green) and honeycomb-like (black) electrolytes obtained by experimental (symbols) and numerical simulations (dashed lines) at 850ºC; current density *(b)* and voltage *(c)* distribution maps obtained by numerical simulation of HC cells operating at V=0.7V in fuel cells conditions. A detail of the distribution map in the cross-section is included.

# TABLES

*Table 1.* Input data calculated for the specific surface area of the TPB in the composite electrodes

| Parameter | Value | Ref. |
|---|---|---|
| Electronic phase volume fraction $\varphi_{el}$ | 0.5 | [41] |
| Ni particle size $d_{el,a}$ | 2.10 μm | |
| 8YSZ particle size in anode $d_{io,a}$ | 2.16 μm | |
| LSM particle size $d_{el,c}$ | 0.54 μm | |
| 8YSZ particle size in cathode $d_{io,c}$ | 0.20 μm | |
| Contact angle $\theta_c$ | 30° | [42] |

***Table 2.*** *Input data introduced in the numerical simulation of the flat SOFCs at T=800ºC*

| Parameter | Value | Units |
|---|---|---|
| Operating temperature | 1073 | K |
| $\Delta p$ anode | 2 | Pa |
| $\Delta p$ cathode | 6 | Pa |
| Exchange current density anode $i_{0a}$ | 0.06 | A/cm$^2$ |
| Exchange current density anode $i_{0c}$ | 0.056 | A/cm$^2$ |
| Specific surface area anode | 6.7 x 10$^5$ | 1/m |
| Specific surface area cathode | 2.1 x 10$^6$ | 1/m |
| Initial cell polarization $V_{pol}$ | 0.05 | V |
| Anode permeability $k_a$ | 6.19 x 10$^{-16}$ | m$^2$ |
| Cathode permeability $k_c$ | 1.29 x 10$^{-16}$ | m$^2$ |
| Equilibrium voltage @ anode $E_{eq,a}$ | 0 | V |
| Equilibrium voltage @ cathode $E_{eq,c}$ | 1.16 | V |
| Electronic conductivity anode $\sigma_a$ | 2 x 10$^6$ | S/m |
| Ionic conductivity anode $\sigma_{s,a}$ | 2.26 | S/m |
| Electronic conductivity cathode $\sigma_c$ | 13600 | S/m |
| Ionic conductivity cathode $\sigma_{s,c}$ | 2.26 | S/m |
| Electrolyte conductivity $\sigma_{ion}$ | 1.1 | S/m |
| Anode porosity $\varepsilon_a$ | 0.23 | - |
| Cathode porosity $\varepsilon_c$ | 0.38 | - |

*Table 3.* Circuit elements value at different temperatures (T=650-850°C). Values are referred to the equivalent circuit of Figure 3. Ohmic resistance ($R_\Omega$), polarization resistances ($R_{P1}$, $R_{P2}$) and constant phase elements ($CPE_1$, $CPE_2$) are obtained.

| T (°C) | $R_\Omega$ (Ω·cm²) | $R_{P1}$ (Ω·cm²) | $CPE_1$ (F/cm²) | $n_1$ | $R_{P2}$ (Ω·cm²) | $CPE_2$ (F/cm²) | $n_2$ |
|---|---|---|---|---|---|---|---|
| 650 | 5.1 | 4.2 | 0.002 | 0.63 | 110 | 0.005 | 0.71 |
| 750 | 2.0 | 1.1 | 0.002 | 0.73 | 11 | 0.009 | 0.68 |
| 850 | 1.3 | 1.1 | 0.011 | 0.63 | 0.7 | 0.035 | 0.81 |